\documentclass[preprint,12pt]{elsarticle}

\usepackage{graphicx}

\usepackage{amssymb}

\journal{Nuclear Physics A}

\begin{document}

\begin{frontmatter}



\title{Initial Conditions and Global Event Properties from Color Glass
Condensate}


\author{Adrian Dumitru}

\address{$^a$Department of Natural Sciences, Baruch College,\\
17 Lexington Avenue, New York, NY 10010, USA\\
$^b$RIKEN BNL Research Center, Brookhaven National
  Laboratory, Upton, NY 11973, USA}

\begin{abstract}
Perturbative unitarization from non-linear effects is thought to
deplete the gluon density for transverse momenta below the saturation
scale. Such effects also modify the distribution of gluons produced in
heavy-ion collisions in transverse impact parameter space.  I discuss
some of the consequences for the initial conditions for hydrodynamic
models of heavy-ion collisions and for hard ``tomographic''
probes. Also, I stress the importance of realistic modelling of the
fluctuations of the valence sources for the small-$x$ fields in the
impact parameter plane. Such models can now be combined with solutions
of running-coupling Balitsky-Kovchegov evolution to obtain controlled
predictions for initial conditions at the LHC.
\end{abstract}

\begin{keyword}
Heavy-ion collisions \sep Gluon saturation \sep Hydrodynamics \sep
Initial Conditions


\end{keyword}

\end{frontmatter}


\section{Introduction}
\label{sec:Intro}

The concept of perturbative saturation of the gluon density of a
hadron at small light-cone momentum $x$ was introduced originally to
preserve unitarity of the scattering amplitude at high
energy~\cite{GLR} which is violated by linear perturbative QCD. In
particular, non-linear processes such as ``gluon recombination''
should prevent a power-law divergence of the (unintegrated) gluon
density $\Phi(x,k_\perp^2)$ at small intrinsic transverse momentum
$k_\perp$, and instead lead to its ``saturation'': $\Phi(x,k_\perp^2)
\sim \log\, 1/k_\perp^2$.

McLerran and Venugopalan suggested that for a large nucleus and at
small $x$ that the problem of gluon saturation could be addressed by
classical methods~\cite{MV}. In their model, the high density $\mu^2$
of valence charges per unit transverse area on the light cone acts as
a source for a classical Yang-Mills field of soft gluons. This
classical field, indeed, was shown to exhibit saturation for
$k_\perp^2 < Q_s^2 \sim \alpha_s^2
\mu^2$~\cite{JalilianMarian:1996xn}. Most important for the present
purposes, however, is that perturbative gluon saturation also predicts
a modified distribution of small-$x$ gluons in transverse impact
parameter space: in the unitarity limit the local density of gluons,
integrated over $k_\perp^2$, is no longer simply proportional to the
density of valence sources as expected in the dilute, linear regime.

The emerging consequences for the initial conditions for hydrodynamic
models of heavy-ion collisions were realized
later~\cite{Hirano:2005xf,Drescher:2006pi}.  Consider a collision of
two heavy ions at non-zero impact parameter. Neglecting fluctuations
of the local density of participants, their overlap area in the
transverse plane has a short axis, parallel to the impact parameter,
and a long axis perpendicular to it. If the produced gluons
equilibrate then the pressure gradients convert this asymmetry of the
initial density profile into a momentum asymmetry called ``elliptic
flow'', $v_2 \sim \langle\cos 2\phi\rangle$. In the absence of any
scales (such as the freeze-out temperature $T_f$, the phase transition
temperature $T_c$, or a non-vanishing mean free path $\lambda$)
hydrodynamics predicts that $v_2$ is proportional to the eccentricity
$\varepsilon$ of the overlap area~\cite{Ollitrault:1992bk},
$\varepsilon = \langle y^{\,2}{-}x^{\,2}\rangle/ \langle
y^{\,2}{+}x^{\,2}\rangle$. The average is taken with respect to the
distribution of produced gluons in the transverse $x$-$y$ plane.

A simple initial condition assumes that by analogy to the Glauber
model for soft particle production $dN/dyd^2r_\perp \sim \rho_{\rm
  part}^{\rm ave}(r_\perp) \equiv (\rho_{\rm part}^A(r_\perp) +
\rho_{\rm part}^B(r_\perp))/2$, where $\rho_{\rm part}^i$ is the
density of participants of nucleus $i$ per unit transverse area. A
$\sim (5-20)\%$ contribution of hard particles needs to be added in
order to fit the centrality dependence of $dN/dy$; their transverse
density scales like $\sim T_{AB}(r_\perp)$.

High-density QCD (the ``Color-Glass Condensate'') predicts a different
distribution of gluons in the transverse plane, corresponding to a
higher eccentricity $\varepsilon$ for intermediate impact
parameters. In particular, when either $A$ or $B$ is dense the number
of produced particles is proportional only to the density of the
dilute collision partner, whose partons add up linearly. Hence, {\em
  in the reaction plane}, $dN/dyd^2r_\perp \sim {\rm
  min}(Q_{s,A}^2,Q_{s,B}^2) \sim {\rm min}(\rho_{\rm part}^A ,
\rho_{\rm part}^B)$ drops more rapidly towards the edge than
$dN/dyd^2r_\perp \sim \rho_{\rm part}^{\rm
  ave}$~\cite{Drescher:2006pi}. Thus, a higher eccentricity is a generic
effect due to a dense target or projectile. Specific numerical
estimates at the limited available RHIC energy do depend on the
model for the unintegrated gluon distribution, however.

\section{Modelling fluctuations of the large-$x$ valence sources}
\label{sec:Large-x}

Numerical estimates for the density distribution of produced partons
require detailed modelling of the fluctuations of the large-$x$
sources in impact parameter space. For peripheral collisions, and for
smaller nuclei such as Cu, this is obvious. However, {\em differences}
of moments of the density distribution, such as the eccentricity
$\varepsilon$, exhibit sensitivity to fluctuations even for central
collisions of heavy nuclei~\cite{Miller:2003kd}. This is due to the
fact that in the presence of fluctuations $\varepsilon\neq 0$ for
central $b\to 0$ collisions (with $x$ and $y$ directions defined via
the principal axes of the particle production zone), while it would
otherwise vanish. The same applies to another moment, the
``triangularity''~\cite{Alver:2010gr}, which gives rise to $\sim
\langle\cos 3\phi\rangle$ contributions to the azimuthal distribution
of particles (a non-vanishing triangularity arises in the CGC
framework away from midrapidity, due to evolution of $Q_s^{A,B}$ with
$y$, even without fluctuations of the sources~\cite{Drescher:2006pi}). 

To model the fluctuations of the valence sources for the small-$x$
fields in the transverse plane one notes that several distinct
transverse distance scales are involved. The radius $R_A$ of a nucleus
is much larger than the radius $R_N$ of a nucleon (the confinement
scale) and one may therefore treat their fluctuations classically; in
other words, we consider a collision of two ``bags'' of nucleons
(quite densely packed, though) instead of using a quantum mechanical
wave function for the nucleus.  The transverse coordinates of the
nucleons can be sampled randomly from a Woods-Saxon
distribution. Multi-particle correlations are usually neglected (see,
however, ref.~\cite{Baym:1995cz}) except for a short-distance hard
core repulsion which enforces a minimal distance $\approx 0.4$~fm
between any two nucleons.

The scale where particle production occurs within the CGC framework is
$\sim 1/Q_s$. It is again much smaller than the radius of a nucleon,
\begin{equation}
\frac{1}{Q_s} \ll R_N \ll R_A ~,
\end{equation}
and so the fluctuations of the valence charge density within a nucleon
could again be treated independently from particle
production. However, in practice nucleons are usually treated as hard
spheres with uniform density.

Once a configuration of valence charges in the transverse plane has
been obtained by Monte-Carlo methods, one constructs the unintegrated
gluon distribution (the small-$x$ fields) $\Phi(x,k_\perp^2;r_\perp)$
at every point $r_\perp$ in the transverse
plane~\cite{Drescher:2007ax}.  The large-$x$ valence charges can still
be treated as frozen sources since their fluctuations in the
transverse plane occur over time scales much larger than the small-$x$
evolution. In the model of Kharzeev, Levin and Nardi~\cite{KLN01}, for
example, the unintegrated ``Weizs\"acker-Williams'' gluon density of a
nucleus (per unit transverse area) is given by
\begin{equation}
\Phi(x,k_\perp^2;r_\perp) \sim \frac{1}{\alpha_s} 
\frac{Q_s^2(x;r_\perp)} {{\rm max}(Q_s^2(x;r_\perp),k_\perp^2)}~.
\end{equation}
$Q_s^2(x;r_\perp)=Q_0^2(r_\perp)\, (x_0/x)^\lambda$ exhibits the
growth of the saturation momentum with energy expected from quantum
evolution. In turn, $Q_0^2(r_\perp)$ corresponds to the density of
valence charge (squared) at the scale $x_0$ and needs to be determined
for each configuration individually.

The fluctuations in the distribution of the hard sources in the
transverse plane should not be confused with their distribution in
color space. In the CGC framework, the color charge density per unit
area of a nucleus or hadron is a stochastic variable, and all
observables need to be averaged over its distribution. In the MV
model, for example, this distribution is a {\em local} Gaussian,
\begin{equation}
W[\rho] \sim \exp\left( - \int d^2r_\perp \frac{{\rm tr}~\rho^2(r_\perp)}
{\mu^2(r_\perp)}\right)~.
\end{equation}
Since there are infinitely many points in a transverse domain on the
order of a {\em fluid cell} (linear dimension $\sim 1/T$) these MV
``fluctuations'' of the valence charge density do not correspond to
fluctuations of the initial condition (particle density) for
hydrodynamics. While quantum evolution in $x$ modifies the MV weight
functional to a non-local distribution, the correlation length
nevertheless remains of order $1/Q_s$ and therefore much smaller than
the size of a fluid cell. Fluctuations of the evolution ladders should
therefore have a small effect on the hydrodynamical evolution and on
``global'' event properties ($v_2$ etc.).

\section{Applications} \label{sec:Applic}
\begin{figure}[t]
  {\includegraphics[width=7cm]{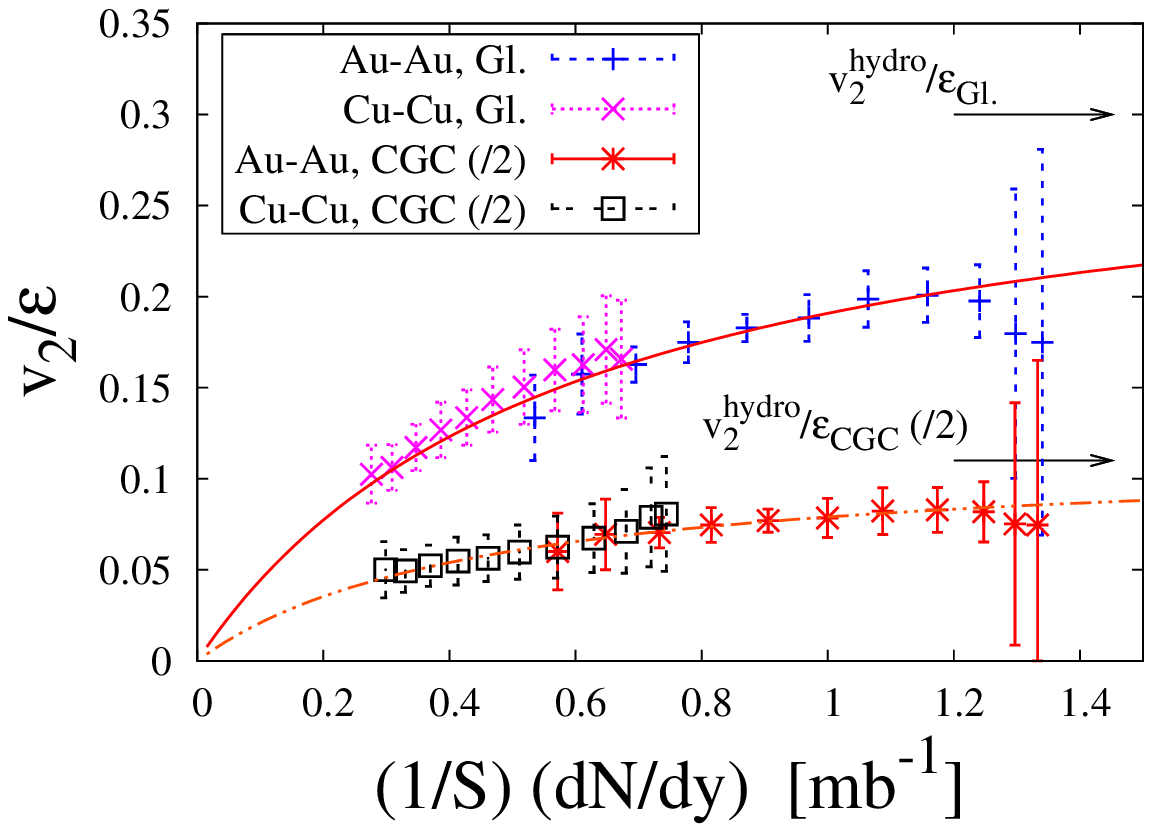}
    \includegraphics[width=6cm]{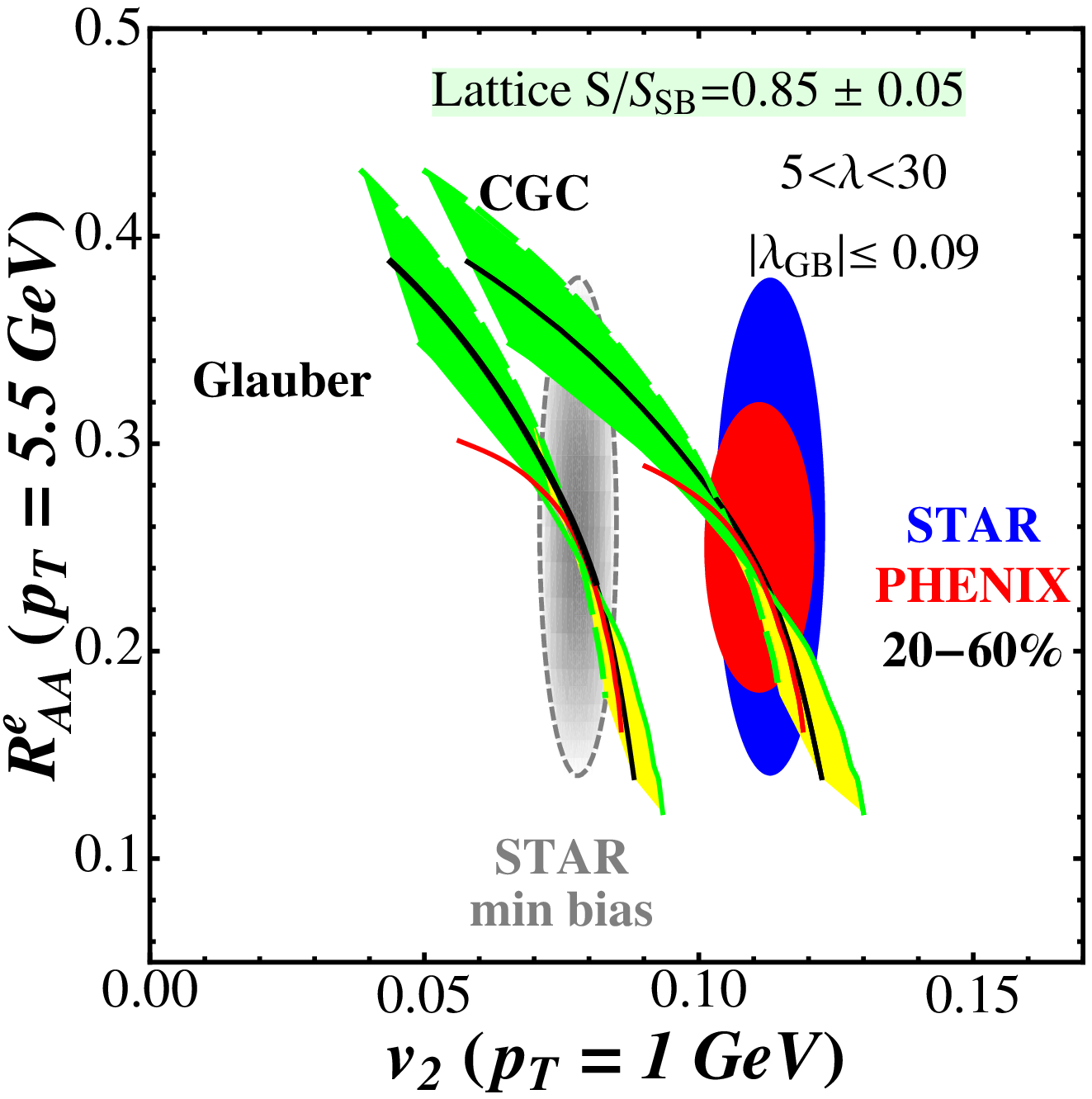}}
    \caption{Left: $v_2/\varepsilon$ versus the transverse
      density~\cite{Drescher:2007cd}; $v_2/\varepsilon_{\rm CGC}$ has
      been scaled by 1/2 for better visibility. Right: Correlation
      between the suppression of hard non-photonic electrons and $v_2$
    of bulk particles expected from a gravity dual model~\cite{Noronha}.}
    \label{fig1}
\end{figure}
Fig.~\ref{fig1} shows the elliptic flow $v_2$ measured in heavy-ion
collisions at RHIC scaled by the eccentricity $\varepsilon$ of the
overlap zone~\cite{Drescher:2007cd}. As already mentioned above, in
the absence of any scales such as a non-zero mean free path,
$v_2/\varepsilon$ would be independent of the transverse density of
particles. Indeed, if the $v_2$ data is scaled by the eccentricity
obtained from a CGC implementation based on the KLN model then the
required breaking of scale invariance is seen to be lower than for
Glauber-like initial conditions. Actual solutions of viscous
hydrodynamics (for $v_2$) appear to confirm this simple observation in
that the {\em slope} of $v_2/\varepsilon$ versus transverse density is
sensitive to the distribution of produced
particles~\cite{ViscoHydro}. State-of-the-art simulations with
realistic QCD equation of state and with initial conditions which
account for the above-mentioned fluctuations will provide further insight.

Other recent studies~\cite{Alver:2010gr,Nexus} have shown that
fluctuations in the initial state which evolve through the
hydrodynamic expansion can give rise to various structures observed in
two-particle angular correlations, such as the ``near-side ridge'' or
the ``away-side cone''. The fact that these correlations are very long
range in rapidity provides a link to CGC physics of heavy-ion
collisions~\cite{CGCcorr}: the classical color fields introduced with
the MV model are boost invariant, while quantum evolution in $\log
1/x$ diminishes correlations only over rapidity intervals exceeding
$\sim 1/\alpha_s$.

The modification of the distribution of produced particles in
$r_\perp$ space in the non-linear regime affects not only the
hydrodynamical evolution but also energy loss of hard probes. This
observable provides a ``tomographic probe'' of the density
distribution of the bulk. The higher eccentricity of CGC-like initial
conditions increases the azimuthal asymmetry of high-$p_\perp$
jets~\cite{Jia:2010ee}. On the other hand, at fixed multiplicity
$dN/dy$ and $v_2$ of bulk particles it reduces the suppression of hard
non-photonic electrons expected from a gravity dual model,
c.f.\ fig.~\ref{fig1} (right) \cite{Noronha}.

\section{Future developments}  \label{sec:Future}

Present CGC based estimates of the initial conditions for
hydrodynamics from fluctuating valence sources usually rely on simple
KLN-like models for the $x$, $k_\perp$ and $r_\perp$ dependence of the
unintegrated gluon distributions (uGD). To improve the reliability of
the approach one should incorporate constraints resulting from the
measured (centrality dependent) $p_\perp$-distributions in d+Au
collisions at central and at forward rapidity. Furthermore, to
actually test the theory of small-$x$ evolution it is important that
extrapolations to LHC energy are based on uGDs which truly solve those
evolution equations. Significant progress has been made recently in
solving the running-coupling Balitsky-Kovchegov equation~\cite{rcBK},
and those results can now be combined with the MC codes which generate
configurations of large-$x$ valence sources~\cite{Drescher:2007ax}.

{\bf Acknowledgments:} This work was supported (in part) by
  the DOE Office of Nuclear Physics through Grant
  No.\ DE-FG02-09ER41620 and by The City University of New York
  through the PSC-CUNY Research Award Program, grants 60060-3940.








\end{document}